\documentclass[reprint,nofootinbib,floatfix,aps,prd]{revtex4-2}

\usepackage{graphicx}
\usepackage{amsmath,amssymb,mathtools}

\usepackage[caption=false]{subfig}

\usepackage{booktabs}
\usepackage{dcolumn}
\usepackage{multirow}
\usepackage{array}
\usepackage{adjustbox}
\usepackage{rotating}

\usepackage{orcidlink}
\usepackage[T1]{fontenc}

\newcommand{\pc}{\,\mathrm{pc}}

\begin{document}

\preprint{000-000-000}

\title{Response to "Late Time Dynamical Dark Energy ... CMB-Distance Ladder Tension"}

\author{Michael S. Turner}
\email{mturner@uchicago.edu}
\affiliation{Kavli Institute for Cosmological Physics,
The University of Chicago, Chicago, IL  60637-1433}
\affiliation{Department of Physics and Astronomy, University of California, Los Angeles, Los Angeles, CA  90095-1547}

\author{Dragan Huterer}
\email{huterer@umich.edu}
\affiliation{Department of Physics and Leinweber Institute for Theoretical Physics,
  University of Michigan, 450 Church St, Ann Arbor, MI 48109, USA}

\date{\today}

\begin{abstract}
We respond to the recent paper by \citet{Efstathiou:2026fax} who makes several statements that appear to be in conflict with our recent paper \cite{Turner:2026zlj}.  We stand by our primary result that direct determinations of $H_0$ can be sensitive to the late-time expansion history.  We also point out where there is significant agreement between his work and ours.

\end{abstract}
\maketitle

\citet{Efstathiou:2026fax} makes two main points in his paper.  First,  that the Hubble tension can be rephrased as a 0.2 mag discrepancy in the calibration of the absolute magnitude of type Ia supernovae (SN) 
as determined by Cepheids and by using the cosmic microwave background (CMB). The bulk of his paper is devoted to this point, which is 
unrelated to the content of our paper.  

His second point is a claim that the Hubble tension is "insensitive to the [...] late time expansion history of the Universe".  This point is related to the content of our paper, but it is not inconsistent with anything we say.  

The primary result of our paper, captured in its title, \textit{The DESI results impact the local determination of} $H_0$, is that the result for $H_0$ that one obtains from low-redshift data ($z\le 0.1$) can depend upon the background cosmology.  

It is straightforward to show why this is.  The apparent magnitude $m$ of an object and its absolute magnitude $M$ are related by
\begin{equation}
\begin{aligned}
    m  = M + 5\log_{\rm 10}\left ( {d_L}/{10\pc}\right ) \nonumber 
    \label{eq:mag}
\end{aligned}
\end{equation}
where $d_L \equiv  (1+z)\int_0^z dx/H(x)$ is the luminosity distance and $H(z)$ is the expansion rate. It is clear that in principle the Hubble constant $H_0$ inferred from measurements of $m$  will depend upon the cosmological model, i.e., $H(z)$.  

For $\Lambda$CDM, $H^2(z) = H_0^2\left[ \Omega_\Lambda + \Omega_M (1+z)^3 \right]$, where $\Omega_\Lambda = 1 -\Omega_M$.  In addition to $H_0$ there is another cosmological parameter, $\Omega_M$.  However, the dependence upon any inference for $H_0$ upon $\Omega_M$ is very slight, cf. Fig.~1 of our paper \cite{Turner:2026zlj}.

For $w_0w_a$CDM, $\Omega_\Lambda$ is replaced by an expression that depends upon $w_0$ and $w_a$. We find that for the models preferred by the DESI BAO measurements the dependence of the inferred value of $H_0$ can be significant, cf. Fig.~2 of our paper \cite{Turner:2026zlj}.  This fact is not disputed in Ref.~\cite{Efstathiou:2026fax}.

The final section of Efstathiou's paper \cite{Efstathiou:2026fax} contains a statement directly relevant to our paper:  "TH26 claim that the evidence for evolving dark energy presented by the DESI collaboration has an impact on the
Hubble tension."  We did not make such a claim.  

Rather, we made two more cautious statements.  In the abstract, we say "The dependence of local determinations of $H_0$ on the background cosmology, ....
is relevant to the Hubble tension."  And in the final sentence of our paper, we state "Resolving the Hubble tension with physics beyond $\Lambda$CDM may involve new physics at both high redshift and low redshift ..."

At the end of section III we also state, "Fixing $\Omega_M$ and assuming that there are no early modifications to the $\Lambda$CDM cosmology, moving along the degeneracy direction of the DESI ellipse for $w_0$ and $w_a$, the CMB value for $H_0$ decreases ... This decrease in the CMB value for $H_0$ tends to maintain the Hubble tension."  This statement is consistent with the Efstathiou's results \cite{Efstathiou:2026fax}.

Our paper was meant to call attention to the fact that the background cosmology can impact local determinations of $H_0$ and thus {\it may be} relevant to the Hubble tension.  We were silent on whether or not it will be important, because to make such a judgment requires a comprehensive alternative description of dark energy.  The $w_0w_a$CDM model is not meant to be such and using it over the large redshift range, $z =0$ to $z = 1100$, could be misleading.
With the exceptions noted above, where the two papers overlap, there is significant agreement.

\bibliographystyle{apsrev4-2}
\bibliography{refs_E26}

@article{Turner:2026zlj,
    author = "Turner, Michael S. and Huterer, Dragan",
    title = "{The DESI results impact the local determination of $H_0$}",
    eprint = "2606.05358",
    archivePrefix = "arXiv",
    journal = "",
    primaryClass = "astro-ph.CO",
    month = "6",
    year = "2026"
}

@article{Efstathiou:2026fax,
    author = "Efstathiou, George",
    title = "{Late Time Dynamical Dark Energy and the CMB-Distance Ladder Tension}",
    eprint = "2607.29562",
    archivePrefix = "arXiv",
    journal = "",
    journal="",
    primaryClass = "astro-ph.CO",
    month = "7",
    year = "2026"
}

\end{document}